\documentclass[aps,prl,twocolumn,groupedaddress,nofootinbib,showpacs]{revtex4}
\usepackage{graphicx}

\begin{document}


\title{Can Dark Matter Annihilation Dominate the Extragalactic
Gamma-Ray Background?}

\author{Shin'ichiro Ando}
\email{ando@utap.phys.s.u-tokyo.ac.jp}
\affiliation{Department of Physics, School of Science, The University of
Tokyo, Tokyo 113-0033, Japan}

\date{Submitted November 22, 2004; accepted February 28, 2005}

\begin{abstract}
Annihilating dark matter (DM) has been discussed as a possible source of
 gamma-rays from the galactic center (GC) and contributing to the
 extragalactic gamma-ray background (EGB).
Assuming universality of the density profile of DM halos, we show that
 it is quite unlikely that DM annihilation is a main constituent of EGB,
 without exceeding the observed gamma-ray flux from the GC.
This argument becomes stronger when we include enhancement of the
 density profiles by supermassive black holes or baryon cooling.
The presence of substructure may loosen the constraint, but only if a
 very large cross section as well as the rather flat profile are
 realized.
\end{abstract}


\maketitle


We have made great progress in our knowledge of what the universe is
composed of.
Surprisingly, we already know that the main constituents of the universe
are not baryonic, but are unknown dark matter (DM) and dark energy.
Recent analyses using observational data of the cosmic microwave
background anisotropy, type Ia supernovae, and large scale structure
precisely give the relic density of these components, $\Omega_{\rm DM} =
0.22$ and $\Omega_\Lambda = 0.73$ \cite{Spergel03}.
In particular for DM, we have some candidates motivated by
particle physics.
The most viable candidate is the lightest supersymmetric particle, which
is a neutralino in most models.
This supersymmetric neutralino can annihilate into final states including
photons via various channels, and these photons might be detectable or
might have already been detected from several astrophysical objects (see
Refs. \cite{Jungman96} for reviews).

In the direction of the galactic center (GC), there are clear gamma-ray
signals in the GeV and TeV energy regions, which have been detected
respectively by the EGRET detector \cite{Mayer-Hasselwander98}, and the
atmospheric \v Cerenkov telescopes (ACTs) such as Whipple
\cite{Kosack04a,Kosack04b}, CANGAROO-II \cite{Tsuchiya04}, and H.E.S.S.
\cite{Aharonian04}.
These gamma-rays from the GC are potentially due to DM annihilation, and
have been extensively studied \cite{Bengtsson90,Fornengo04}.
These results show that if the density profile of the galactic central
region is cuspy enough, as suggested by $N$-body simulations such as by
Navarro, Frenk, and White \cite{Navarro96} (hereafter NFW) and Moore et
al. \cite{Moore99} (hereafter M99), then the gamma-ray fluxes can be
explained by the neutralino annihilation with a cross section that gives
the proper relic density $\Omega_{\rm DM}$.

On the other hand, analyses of the diffuse EGRET emission shows the
signature of an extragalactic gamma-ray background (EGB) in the GeV
range \cite{Sreekumar98,Strong04}.
Annihilating DM may also significantly contribute to this EGB flux.
Using the hierarchical clustering formalism that is now widely accepted,
several authors gave the flux predictions, investigating the effect of
DM clustering or substructure
\cite{Ullio02,Taylor03,Bergstrom01a,Elsasser04}, and suggested that the
EGB data can be explained well by including the DM component.
In particular, a bump around 3 GeV, discovered by the recent
reanalysis \cite{Strong04}, may be the signature \cite{Elsasser04}.

In this Letter, we investigate the gamma-ray signature from the GC and
EGB together, assuming that the halo profile is universal as suggested
by recent $N$-body simulations \cite{Navarro04}.
With this self-contained treatment, we point out that the annihilating
DM cannot be a main constituent of the observed EGB without exceeding
observational bounds imposed by gamma-ray measurements of the GC.
Since both the GC and EGB fluxes should be predicted using the same
cross section and mass of the DM particle, they are connected if we
specify these ingredients.
We also show that the main conclusion of this Letter has a quite robust
characteristic such that it does not depend on uncertainties concerning
both the particle physics models and other astrophysical inputs.
The latter includes the central spike of halos due to presence of
supermassive black holes (SMBHs) \cite{Gondolo99} or to the baryon
cooling \cite{Gnedin04}, and the significant enhancement of the EGB flux
due to the inclusion of halo substructure.

{\it Gamma-rays from the galactic center.}---%
The number flux of high-energy gamma-rays due to DM annihilation can be
calculated with the following formulation:
\begin{eqnarray}
\Phi_\gamma^{\rm GC} (E_\gamma)\Delta\Omega
 &=& 9.4\times 10^{-11}~\mathrm{cm^{-2}~s^{-1}}~m_{\chi,2}^{-2}
 \nonumber\\ &&{}\times
 \langle\sigma v\rangle_{-26}
 \frac{dN_\gamma}{dE_\gamma}\overline{J(\Delta\Omega)}\Delta\Omega,
  \label{eq:GC flux}
\end{eqnarray}
where $m_\chi = 100 m_{\chi,2}$ GeV is the mass of the DM particle,
$\langle\sigma v\rangle = 10^{-26} \langle\sigma v\rangle_{-26}$ cm$^3$
s$^{-1}$ is average value of the annihilation cross section times the
relative velocity (assumed to be independent of $v$), and $dN_\gamma
/dE_\gamma$ represents the gamma-ray spectrum per annihilation, for
which we use a simple parameterization as $dN_\gamma /dE_\gamma \simeq
(0.73/m_\chi)e^{-7.76E_\gamma / m_\chi}/ [(E_\gamma / m_\chi)^{1.5}
+0.00014]$ \cite{Bergstrom01a}.
(Although this parameterization may be less precise, it is sufficient
for our purpose.)
$\overline{J(\Delta\Omega)}$ represents the average value of the
following quantity:
\begin{equation}
J(\psi)=\frac{1}{\mathrm{8.5~kpc}}
 \int_{\rm l.o.s.}dl(\psi)
 \left(\frac{\rho (r(\psi,l))}{\mathrm{0.3~GeV~cm^{-3}}}\right)^2,
  \label{eq:J}
\end{equation}
over the detector angular resolution $\Delta\Omega$ [$\Delta\Omega
\simeq 2\times 10^{-3}~(4\times 10^{-5})$ for EGRET (ACTs)], and
normalized to the local value.
The integration is performed along the line of sight (l.o.s.) labeled
by angle deviation $\psi$ from the GC ($\psi=0$ for the direction to
the GC), and $r$ is the direction to the integrated point from the GC.

For the density profile of the DM halos $\rho (r)$, we use the NFW and
M99 models, which are characterized by the central slopes of $\gamma =
1$ and 1.5, respectively, where $\gamma$ is defined by $\rho(r)\propto
r^{-\gamma}$ for small radii.
While the most recent $N$-body simulations suggest there is no
asymptotic slope and do not give such a cuspy profile as M99 (NFW may
still be marginally consistent) \cite{Navarro04}, we use these two
profiles as our reference models, in order to investigate how our
conclusion changes with the selected profile.
Furthermore, considering some other physical processes, it would still
be possible to obtain significant enhancement of the central density.
The proposed mechanism giving such a ``spike'' that may be steeper
than the M99 profile is the accretion of DM particles onto a central
SMBH \cite{Gondolo99} or the effect of baryon cooling \cite{Gnedin04}.

\begin{table}
\caption{Angular acceptance $\overline{J(\Delta\Omega)}$ of the GC
 gamma-rays, and local values of enhancement factor $f(0)$ for the
 EGB flux. \label{table:J_average}}
\begin{tabular}{cccc}\hline\hline
 Model & $\overline{J(2\times 10^{-3})}$ & $\overline{J(4\times
 10^{-5})}$ & $f(z=0)$\\
 \hline
 NFW & $7\times 10^2$ & $5\times 10^3$ & $2\times 10^4$\\
 M99 & $6\times 10^4$ & $4\times 10^6$ & $2\times 10^5$\\
 \hline\hline
\end{tabular}
\end{table}

In Table \ref{table:J_average}, we summarize the values of
$\overline{J(\Delta\Omega)}$ evaluated with the NFW and M99 density
profiles.
Because of its steeper profile in the central region, the M99 profile
gives much larger values of $\overline{J(\Delta\Omega)}$ than NFW.
This difference becomes more prominent when we use detectors with
better resolution, since a more concentrated region can be probed.
In Table \ref{table:J_average}, we used the cutoff scale $10^{-8}$ kpc,
and the profile was assumed to be flat within that radius.
This is because without this implementation, the l.o.s. integration
would diverge mathematically at very small radius.
The cutoff scale may be physically determined by the annihilation
itself, scattering of DM particle off stars, contraction of baryons,
or the presence of a central SMBH.
Because even the most recent simulations do not reach the very inner
region of the halo (but as large as $\sim 0.1$--1 kpc), the choice of
the cutoff scale is a nontrivial problem.
However, the GC flux is rather weakly dependent on this parameter in
the reasonable range \cite{Fornengo04}, and the uncertainty does not
strongly affect our conclusion.

\begin{figure}
\includegraphics[width=8.5cm]{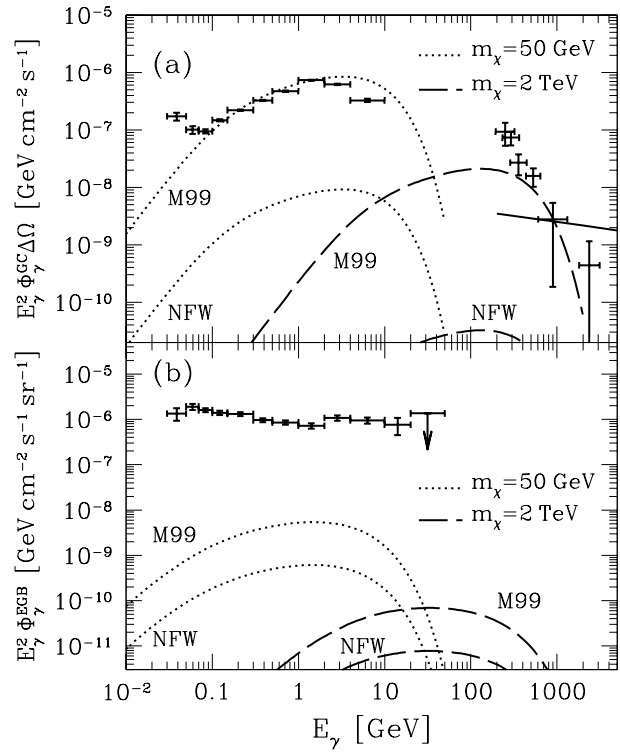}
\vspace{-0.3cm}
\caption{(a) Gamma-ray flux from the GC from annihilating DM, with mass
 50 GeV or 2 TeV, evaluated with the NFW and M99 profiles.
 Data from EGRET \cite{Mayer-Hasselwander98} and CANGAROO-II
 \cite{Tsuchiya04} are also plotted; the H.E.S.S. result
 \cite{Aharonian04} is shown as a solid line. (b) EGB intensity from DM
 annihilation. EGRET data points \cite{Strong04} are also
 plotted. \label{fig:flux}}
\vspace{-0.3cm}
\end{figure}

Figure \ref{fig:flux}(a) shows the gamma-ray flux from the GC due to
annihilating DM, with masses 50 GeV or 2 TeV.
In deriving these expressions, we assumed $\langle\sigma v\rangle_{-26}
=3$ that is considered to be appropriate for leaving the observed relic
density of DM \cite{Jungman96}; larger values than this would imply a
lower relic density, requiring an additional DM component.
Data points in 0.03--10 GeV and 0.2--2 TeV are taken from the EGRET
\cite{Mayer-Hasselwander98} and CANGAROO-II \cite{Tsuchiya04} papers.
A solid line above 2 TeV represents the power-law fit to the H.E.S.S.
data \cite{Aharonian04} (the recent Whipple result is consistent with
the H.E.S.S. data \cite{Kosack04b}).
Correspondingly, theoretical curves are evaluated using Eq. (\ref{eq:GC
flux}) with $\Delta\Omega =2\times 10^{-3}$ for 50 GeV, and with
$\Delta\Omega =4\times 10^{-5}$ for 2 TeV DM particles.
We can clearly see that, in the case of $m_\chi = 50$ GeV, the flux
evaluated with the M99 profile can easily be quite consistent with the
EGRET data points over the wide range of energy.
With the NFW profile, on the other hand, we predict considerably less
flux.
TeV gamma-rays may also be dominated by DM component, although both the
flux and spectral shape are still controversial.

{\it Extragalactic gamma-ray background.---}%
The EGB flux estimation involves somewhat more information, e.g., that
on the cosmological clustering of DM halos.
The intensity of EGB is calculated by
\begin{eqnarray}
\Phi_\gamma^{\rm EGB} (E_\gamma) &=&\frac{c}{4\pi H_0}
 \frac{\langle\sigma v\rangle}{2}
 \frac{\Omega_\chi^2\rho_{\rm crit}^2}{m_\chi^2}
 \int dz \frac{(1+z)^3}{h(z)}\nonumber\\
 &&{}\times\frac{dN_\gamma(E_\gamma^\prime)}{dE_\gamma^\prime}
  f(z) e^{-\tau (z,E_\gamma)},
  \label{eq:EGB intensity}
\end{eqnarray}
where $E_\gamma^\prime = (1+z) E_\gamma$, $h(z) = [(1+z)^3\Omega_m +
\Omega_\Lambda]^{1/2}$ and $\rho_{\rm crit}$ is the critical density.
We also include the effect of gamma-ray absorption by $e^{-\tau}$,
which is caused by pair annihilation with the diffuse extragalactic
background light in the infrared or optical wavebands \cite{Salamon98}.
This effect changes the EGB flux at TeV regions, but its extent is
too small to affect the results.
Hierarchical clustering of the DM halos is included in an intensity
multiplier $f(z)$ \cite{Ullio02,Taylor03}.
For evaluating it we used the halo mass function based on the
ellipsoidal collapse model \cite{Sheth99} with a lower mass cutoff of
$M_{\rm min}=10^6 M_\odot$, which may be determined by the validity of
hierarchical clustering formalism, self-limitation due to annihilation
itself, or nuclear and star formation activities \cite{Taylor03}.
Varying it over a reasonable range ($10^4$--$10^8M_\odot$) changes the
EGB flux only by a factor of 2 or less \cite{Ullio02,Taylor03}.
To evaluate a concentration parameter that represents how the bulk of
mass in each halo concentrates in the central region, we used a publicly
available numerical code by \citet*{Eke01}.
The resulting values of $f(0)$ for each profile are summarized in the
fourth column of Table \ref{table:J_average}, and we note that our
result is consistent with that of Refs. \cite{Ullio02,Taylor03}.
We should note that these values are significantly smaller than those
adopted by previous studies such as Refs. \cite{Elsasser04}.
This large discrepancy potentially comes from uncertainty concerning the
concentration parameter, and the presence or absence of substructures.
We also discuss these possibilities later.

We show in Fig. \ref{fig:flux}(b) the EGB intensity, with the same
physical inputs as in the GC flux calculation.
This shows that without any processes that give much larger $f(z)$, the
expected DM contribution to the EGB flux is considerably smaller than
the observed value.
We also note that the dependence on the adopted profile is less
prominent in the case of EGB, compared with strong dependence of
gamma-ray flux from the GC.
This is because the EGB flux is less sensitive to the very central
region of the halo.
The weaker dependence of $f(0)$ on the profile shown in Table
\ref{table:J_average} also reflects the same characteristic.
With our canonical model, it is much more difficult to explain the
observed EGB intensity mainly by annihilating DM component than
gamma-ray flux from the GC; it requires an additional boost by more
than two orders of magnitude.

{\it Constraints on annihilating dark matter component.}---%
The contribution of DM annihilation to the EGB flux is quite strongly
constrained by the GC gamma-ray observations, and this result is rather
robust, independent of uncertainties in the particle physics models.
As a first step, we introduce a boost factor $b$ for both the GC
gamma-rays and EGB, as a correction to the canonical 
predictions of each flux.
All the corrections due to the other astrophysical and particle physical
possibilities are included in $b$.
In order for the DM annihilation to be a main constituent of the
observed fluxes, the required values of $b$ should be very close to the
following quantity: $b^{\rm max}\equiv\min_{i}\left[\Phi_{\gamma,i}^{\rm
obs}/\Phi_{\gamma}^{\rm th}(E_{\gamma,i})\right]$, where $i$ represents
bin-number of each observation, and $\Phi_{\gamma,i}^{\rm obs}$ and
$\Phi_\gamma^{\rm th}$ are the observed intensity in $i$th bin and
theoretical prediction given by Eqs. (\ref{eq:GC flux}) and
(\ref{eq:EGB intensity}), respectively.
By taking minimum over all the bins, we renormalize the flux with
keeping its shape, so as not to exceed the data points, which should be
regarded as rigorous upper limits.

\begin{figure}
\includegraphics[width=8.5cm]{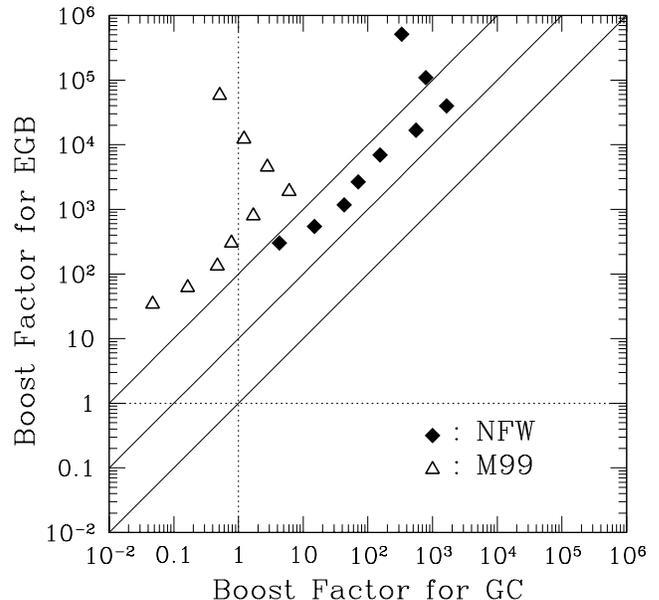}
\vspace{-0.2cm}
\caption{Boost factors, $b_{\rm GC}^{\rm max}$ and $b_{\rm EGB}^{\rm
 max}$, required to make DM annihilation a main component of the data,
 for the NFW and M99 profiles. Assumed masses are 10, 20, 50,
 100, 200, 500, 1000, 2000, and 5000 GeV (from bottom to top). The solid
 and dotted lines represent $b_{\rm EGB}=b_{\rm GC}, 10 b_{\rm GC}, 100
 b_{\rm GC}$ and $b_{\rm GC} = 1, ~ b_{\rm EGB} = 1$, plotted for
 comparison. \label{fig:boost}}
\vspace{-0.5cm}
\end{figure}

From Fig. \ref{fig:flux}, we can confirm that the value of $b$ required
for the GC gamma-ray data is much smaller than that for EGB, i.e.,
$b_{\rm GC}^{\rm max} \ll b_{\rm EGB}^{\rm max}$.
These required values ($b_{\rm GC}^{\rm max}, b_{\rm EGB}^{\rm max}$)
are plotted in Fig. \ref{fig:boost}, for both the NFW and M99 profiles
and for several assumed DM particle masses.
We used the CANGAROO-II data in the TeV region, since this would be
more conservative for our purpose.
If we add corrections that are related only to particle physics models
(especially by renormalizing the cross section), we obtain the relation
between two boost factors as $b_{\rm EGB}=b_{\rm GC}$, since the
correction is common to the both cases.
In this case, because of the relation $b_{\rm EGB}^{\rm max} \gg
b_{\rm GC}^{\rm max}$ for all the models as shown in Fig.
\ref{fig:boost}, we cannot explain the EGB data mainly by annihilating
DM.
Otherwise, it would overproduce gamma-rays from the GC compared to the
data.
In addition, Fig. \ref{fig:boost} also shows that this tendency is
more prominent for the M99 profile than NFW.
Although we restrict our argument within these two specific profiles,
the conclusion derived here is general and applicable to other profile
choices, as discussed below for more specific examples.

We note that the strong gamma-ray signal may not be coming from
the GC; it has been suggested that the EGRET GeV source position is
offset from the dynamical center of the galaxy, Sgr A$^\ast$, at
roughly 95\% C.L. \cite{Hooper02}.
If this is true, it also strengthens our argument, because it suggests
that the most of the gamma-rays come from other astrophysical sources,
and DM component should be significantly smaller than the EGRET data.

{\it Other astrophysical possibilities.---}%
A SMBH, due to its deep potential well, could accrete a significant
amount of DM particles, and this would make the density spike in the
central region of halos \cite{Gondolo99}.
It has also been pointed out that the infall of baryons due to
radiative cooling could lead the DM compression in the GC
\cite{Gnedin04}.
Both these effects, potentially and significantly, enhance gamma-ray
signals from DM halos.
It should be noted that an enhancement of the central density profile
by any possible processes strengthens our main conclusion.
This is because the GC gamma-ray flux is much more sensitive to the
slope of the central region, and the resulting relation of the boost
factors, $b_{\rm GC} > b_{\rm EGB}$, prevents the DM component from
becoming dominant in EGB, without violating the gamma-ray observations
of the GC.

On the other hand, if the density profile in the central region of
halos is less steep than the NFW (due to, e.g., rather large inner
cutoff radius, as already mentioned), the required relation between the
boost factors would become close to $b_{\rm EGB}^{\rm max} \sim
b_{\rm GC}^{\rm max}$.
In this case, however, we should note that the DM component becomes
{\it absolutely} difficult to be dominant both in the GC and EGB, while
it is {\it relatively} easier to explain the EGB flux without
overproducing the GC gamma-rays; it requires, e.g., much larger cross
section, which is unlikely.
For example, the calculation using the profile with $\gamma = 0.5$ and
$m_\chi = 100$ GeV gives $b_{\rm EGB}^{\rm max} \simeq 5 b_{\rm GC}
^{\rm max} = 3 \times 10^3$.

Recent $N$-body simulations suggest the presence of DM substructure,
although it is not observationally confirmed.
According to the literature, this might boost the GC gamma-ray and EGB
flux by at most a factor of a few \cite{Stoehr03} and about an order
of magnitude \cite{Ullio02,Taylor03}, respectively.
Therefore, we obtain the relation, $b_{\rm EGB} \alt 10 b_{\rm GC}$
that is still below the required points shown in Fig. \ref{fig:boost}.
It suggests that even the inclusion of substructure cannot provide a
way that violates the main thrust in this Letter.
In the previous studies of the EGB flux \cite{Elsasser04}, the
intensity multiplier as large as $f(0)\simeq 10^7$ (for the M99
profile) was used, which is about a factor of 50 larger than our value
(see Table \ref{table:J_average}).
This discrepancy may come from the different choice of the
concentration parameter, in addition to inclusion of substructure.
The former is extensively discussed in Ref. \cite{Ullio02}, and found
to give an uncertainty of a factor $\sim 5$.
Even if we use this extreme boost factor for the EGB ($b_{\rm EGB}
\sim 50 b_{\rm GC}$), for the M99 profile it still requires some
additional effects that enhances the EGB flux with changing the GC
gamma-rays by a significantly smaller amount.
For the NFW or less steep profile, while it might provide a solution to
the relative smallness of the predicted EGB flux, we still require
considerable amount of corrections, which is physically unlikely.


\begin{acknowledgments}
The author is grateful to the anonymous referees for valuable comments,
and also thank John Beacom for useful discussions.
This work was supported by a Grant-in-Aid for JSPS Fellows.
\end{acknowledgments}

\vspace{-0.3cm}

\bibliography{refs}

\end{document}